\documentstyle[12pt]{article}

\def\Re{{\cal R \mskip-4mu \lower.1ex \hbox{\it e}\,}}
\def\Im{{\cal I \mskip-5mu \lower.1ex \hbox{\it m}\,}}
\def\ie{{\it i.e.}}

\def\etal{{\it et al.}}
\def\ibid{{\it ibid}.}
\def\sub#1{_{\lower.25ex\hbox{$\scriptstyle#1$}}}

\def\to{\rightarrow}

\def\subw{_{\rm w}}
\def\mh{\ifmmode m\sbl H \else $m\sbl H$\fi}
\def\mch{\ifmmode m_{H^\pm} \else $m_{H^\pm}$\fi}
\def\mt{\ifmmode m_t\else $m_t$\fi}
\def\mc{\ifmmode m_c\else $m_c$\fi}
\def\mz{\ifmmode M_Z\else $M_Z$\fi}
\def\mw{\ifmmode M_W\else $M_W$\fi}
\def\mws{\ifmmode M_W^2 \else $M_W^2$\fi}
\def\mhs{\ifmmode m_H^2 \else $m_H^2$\fi}   
\def\mzs{\ifmmode M_Z^2 \else $M_Z^2$\fi}
\def\mts{\ifmmode m_t^2 \else $m_t^2$\fi}
\def\mcs{\ifmmode m_c^2 \else $m_c^2$\fi}
\def\mchs{\ifmmode m_{H^\pm}^2 \else $m_{H^\pm}^2$\fi}
\def\ztwo{\ifmmode Z_2\else $Z_2$\fi}
\def\zone{\ifmmode Z_1\else $Z_1$\fi}
\def\mtwo{\ifmmode M_2\else $M_2$\fi}
\def\mone{\ifmmode M_1\else $M_1$\fi}
\def\tb{\ifmmode \tan\beta \else $\tan\beta$\fi}
\def\xw{\ifmmode x\subw\else $x\subw$\fi}
\def\ch{\ifmmode H^\pm \else $H^\pm$\fi}
\def\lum{\ifmmode {\cal L}\else ${\cal L}$\fi}
\def\inpb{\ifmmode {\rm pb}^{-1}\else ${\rm pb}^{-1}$\fi}
\def\infb{\ifmmode {\rm fb}^{-1}\else ${\rm fb}^{-1}$\fi}
\def\epem{\ifmmode e^+e^-\else $e^+e^-$\fi}
\def\ppb{\ifmmode \bar pp\else $\bar pp$\fi}
\def\mpl{\ifmmode \overline M_{Pl}\else $\bar M_{Pl}$\fi}

\newskip\zatskip \zatskip=0pt plus0pt minus0pt
\def\matth{\mathsurround=0pt}
\def\lsim{\mathrel{\mathpalette\atversim<}}
\def\gsim{\mathrel{\mathpalette\atversim>}}
\def\atversim#1#2{\lower0.7ex\vbox{\baselineskip\zatskip\lineskip\zatskip
  \lineskiplimit 0pt\ialign{$\matth#1\hfil##\hfil$\crcr#2\crcr\sim\crcr}}}

\def\be{\begin{equation}}
\def\ee{\end{equation}}
\def\bea{\begin{eqnarray}}
\def\eea{\end{eqnarray}}
\renewcommand{\thefootnote}{\fnsymbol{footnote}}

\begin{document} \begin{titlepage} 
\rightline{\vbox{\halign{&#\hfil\cr
&SLAC-PUB-8298\cr
&November 1999\cr}}}
\vspace{1in} 
\begin{center}

{\Large\bf  Bulk Gauge Fields in the Randall-Sundrum Model}
\footnote{Work supported by the Department of 
Energy, Contract DE-AC03-76SF00515}
\medskip

\normalsize 
{\large H. Davoudiasl, J.L. Hewett and T.G. Rizzo \\}
\vskip .3cm
Stanford Linear Accelerator Center \\
Stanford CA 94309, USA\\
\vskip .3cm

\end{center}

\begin{abstract} 

We explore the consequences of placing the Standard Model gauge fields 
in the bulk of the recently proposed localized gravity model of Randall and 
Sundrum.  We find that the Kaluza Klein excitations of these fields are
necessarily strongly coupled and we demonstrate that current precision 
electroweak data constrain the lowest states to lie above 
$\simeq 23$ TeV.  Taking the weak scale to be $\sim 1$ TeV,
the resulting implications on the model parameters force
the bulk curvature, $R_5$, to be larger than the higher dimensional Planck 
scale, $M$, violating the consistency of the theory.  In turn, 
to preserve $|R_5|\lsim M^2$, the weak
scale must be pushed to $\gsim 100$ TeV. 
Hence we conclude that it is
disfavored to place the Standard Model gauge fields in the bulk of this 
model as it is presently formulated. 

\end{abstract}

\renewcommand{\thefootnote}{\arabic{footnote}} \end{titlepage} 


\section{Introduction}

The possibility of extra space-like dimensions with accessible physics near 
the TeV scale{\cite {old}} has opened a new avenue for explaining 
the gauge hierarchy.  The models which address the hierarchy 
make use of our ignorance about gravity, in particular, the fact that gravity 
has yet to be probed at energy scales much above $10^{-3}$ eV in laboratory 
experiments. The prototype scenario in this class of theories is due to 
Arkani-Hamed, Dimopoulos and Dvali{\cite {nima}} who use the volume 
associated with large extra dimensions, which may be as sizable as a fraction 
of a millimeter, to bring the $D$-dimensional Planck scale down to 
a few TeV.  Here, the gauge hierarchy problem is recast into the issue of
stabilizing the rather large ratio between the TeV Planck scale and the 
compactification scale of the extra dimensions.  Nonetheless, the 
phenomenological{\cite {pheno}} and astrophysical{\cite {astro}} 
implications of this model have been examined by a large number of authors.

More recently, Randall and Sundrum (RS){\cite {rs1}} have proposed an 
alternative scenario wherein the hierarchy is generated by an
exponential function of the compactification radius, called a warp factor.
Unlike the model of Arkani-Hamed \etal, they 
assume a 5-dimensional non-factorizable geometry, based on a slice
of $AdS_5$ spacetime.  Two 3-branes, one being `visible' with the other being
`hidden', with opposite tensions rigidly reside at
$S_1/Z_2$ orbifold fixed points, taken to be $\phi=0,\pi$, where $\phi$ is
the angular coordinate parameterizing the extra dimension.  It is assumed that 
the extra-dimensional bulk is only populated by gravity, and that the SM lies 
on the brane with negative tension at $\phi=\pi$.   
Gravity is localized on the Planck brane at $\phi=0$.  The 
solution to Einstein's equations for this configuration, maintaining
4-dimensional Poincare invariance, is given by the 5-dimensional metric
\be
ds^2=e^{-2\sigma(\phi)}\eta_{\mu\nu}dx^\mu dx^\nu+r_c^2d\phi^2 \,,
\label{5metric}
\ee
where the Greek indices run over ordinary 4-dimensional spacetime, 
$\sigma(\phi)=kr_c|\phi|$ with $r_c$ being the compactification radius of the
extra dimension, and $0\leq |\phi|\leq\pi$.  Here $k$ is a scale of
order the Planck scale and relates the 5-dimensional Planck scale $M$ to the 
bulk cosmological constant.  Similar configurations have also been found to 
arise in M/string-theory\cite{strings}.  Here, it is assumed that the
5-dimensional curvature $R_5$, where $R_5=-20k^2$, satisfies $|R_5|<M^2$ with
$M\sim\mpl$ (where $\mpl \simeq 2.44 \times 10^{18}$ GeV is the reduced 
Planck mass) so that this solution for the bulk metric can be 
trusted\cite{rs1}.  Otherwise, 
higher order terms in the curvature would need to be kept in the initial 
action to maintain self-consistency. 

Examination of the action in the 
4-dimensional effective theory in the RS scenario yields\cite{rs1}
\be
\mpl^2={M^3\over k}(1-e^{-2kr_c\pi}) 
\label{mpl}
\ee
for the reduced effective 4-D Planck scale.  
A field on the SM brane with the fundamental mass
parameter $m_0$ will appear to have the physical mass $m=e^{-kr_c\pi}m_0$.
TeV scales are thus generated from fundamental scales of order $\mpl$
via a geometrical exponential factor and the observed scale hierarchy is
reproduced if $kr_c\simeq 11-12$.  Due to the exponential nature of the
warp factor, no additional large hierarchies are generated.  In fact, it has 
been demonstrated\cite{gw2} that the magnitude
of $1/r_c$ in this scenario can be stabilized without any fine tuning of 
parameters.  This model thus provides an interesting interpretation of the 
electroweak scale.

In our recent analysis{\cite {dhr}}, we examined the phenomenological 
implications and constraints on the RS model that arise from the direct
resonant production and exchange of weak scale Kaluza-Klein (KK) towers of 
gravitons.  In this work we consider adding the SM gauge 
fields to the RS bulk under the assumption that they make 
little contribution to the bulk energy density so that the solution of 
Einstein's equations remains valid, \ie, the stress energy 
tensor due to SM gauge fields in the bulk
is far smaller than the size of the bulk 
cosmological constant. The possibility that the SM gauge fields may appear in 
the bulk of models with flat, factorizable geometries has been examined in 
detail{\cite {old,sminbulk,getV}} for a wide variety of 
reasons, including the attainment of
low energy coupling constant unification{\cite {guts}}. 
Here, we will demonstrate that the spectra and couplings of the bulk gauge
field KK towers are qualitatively different in the RS model
of localized gravity than in the case with factorizable geometry.  In addition,
we will show that the
resulting phenomenological constraints on the model parameters
lead to a potential internal inconsistency within the theory and thus gauge
fields cannot exist in the bulk without some modification to the theory.

We remind the reader that in the case with a factorizable metric and
one extra dimension compactified on $S^1/Z_2$, ($i$) the masses of the 
KK excitations are equally spaced, given simply by the relation 
$m_n=n/R$, with $R$ being the compactification radius, ($ii$) the SM 
chiral fermions are assumed to naturally remain on the SM brane at the 
orbifold fixed point since they live in the ``twisted'' 
sector of string theory, and ($iii$) the ratio of the couplings to wall 
fermions of the excited KK states to that of the zero mode is simply $\sqrt 2$ 
for all $n$.  While below we retain the second assumption, we will see that
the other results will be quite different in the RS model.
We also note that we do not 
need to specify whether the Higgs scalar is also a bulk field, but if it
does reside in the bulk, it must be $Z_2$-even in order to obtain the zero-mode
Higgs on the SM brane.
In the remainder of the paper we first derive the KK spectrum of the gauge
fields and their couplings to fermions, and then examine the phenomenological
consequences of their contributions to electroweak radiative corrections.
We summarize our results and their implications on the theory in the 
conclusions.

\section{The Gauge Field KK Spectrum}

In what follows we derive the KK spectrum of a $U(1)$ bulk 
gauge field $A_{_M}$ (where the upper case Roman indices extend 
over all 5 dimensions) in the effective 4-dimensional
theory.  The extension to the case of non-Abelian fields is straightforward.
Here, we assume that the $A_\mu$ (where the Greek indices run over ordinary
4-dimensional spacetime) are $Z_2$-even and that $A_4$ is $Z_2$-odd with 
respect to the extra dimension $x^4$.  This choice of $Z_2$ parity preserves 
the gauge-fermion interactions and ensures that $A_4$ does not have a
zero mode in the effective 4-dimensional 
theory.  The 5-dimensional action $S_{_A}$ for a pure $U(1)$ gauge theory
is given by
\begin{equation} 
S_{_A} = - \frac{1}{4} \int d^5 x \, \sqrt{- G} \, \, 
G^{^{M K}} G^{^{NL}} F_{_{KL}} F_{_{MN}}\,,
\label{SA}
\end{equation}
where $\sqrt{- G} \equiv |det \left(G_{_{M N}}\right)|^{1/2} 
= e^{- 4 \sigma}$ and 
$F_{_{MN}}$ is the 5-dimensional field strength tensor given by 
\begin{equation}
F_{_{MN}} = \partial_{_M} A_{_N} - \partial_{_N} A_{_M}\,. 
\label{FMN}
\end{equation}
Note that this definition does not involve the affine 
connection terms due to the antisymmetry of $F_{_{MN}}$.  
After an integration by parts, Eq. (\ref{SA}) yields
\begin{equation}
S_{_A} = - \frac{1}{4} \int d^5 x \left[\eta^{\mu \kappa} 
\eta^{\nu \lambda} F_{\kappa \lambda} F_{\mu \nu} 
- 2 \, \eta^{\nu \lambda} A_\lambda \, \partial_4 
\left(e^{- 2 \sigma} \partial_4 A_\nu \right) \right]\,,
\label{SAINT}
\end{equation}
where we have used gauge freedom to choose $A_4 = 0$.  This is
consistent with the gauge invariant equation 
$\oint d x^4 A_4 = 0$, which results from our assumption 
that $A_4$ is a $Z_2$-odd function of the extra dimension.  
This choice eliminates $A_4$ from the theory on the 3-brane, 
but it will not disturb the gauge invariance of the action in the effective
4-dimensional theory, as we will see below.

Let the KK expansion of $A_\mu$ be given by 
\begin{equation} 
A_{\mu} (x, \phi) = \sum_{n = 0}^\infty A_{\mu}^{(n)} (x) \, 
\frac{\chi^{(n)}(\phi)}{\sqrt{r_c}}\,,
\label{Amu}
\end{equation}
with $x^4 = r_c \phi$.  Using this expansion in Eq. (\ref{SAINT}) 
and integrating over $\phi$ gives
\begin{equation}
S_{_A} = \int d^4 x \sum_{n = 0}^\infty \left[- \frac{1}{4} \, 
\eta^{\mu \kappa} \eta^{\nu \lambda} F^{(n)}_{\kappa \lambda} 
F^{(n)}_{\mu \nu} - \frac{1}{2} \, m_n^2 \eta^{\nu \lambda} 
A^{(n)}_\nu A^{(n)}_\lambda \right]\,,
\label{S4D}
\end{equation}
where $F^{(n)}_{\mu \nu} = \partial_\mu A^{(n)}_\nu - 
\partial_\nu A^{(n)}_\mu$, and we have required that the $\phi$-dependent
wavefunctions satisfy the orthonormality condition
\begin{equation}
\int_{- \pi}^{\pi} d \phi \, \chi^{(m)} \chi^{(n)} = \delta^{m n} 
\label{Norm}
\end{equation}
and the differential equation
\begin{equation}
\frac{- 1}{r_c^2} \frac{d}{d \phi} \left(e^{- 2 \sigma} 
\frac{d}{d \phi} \, \chi^{(n)}\right) = m_n^2 \, \chi^{(n)}\,.
\label{Eigen}
\end{equation}
The expression in Eq. (\ref{S4D}) is the action for gauge 
fields $A^{(n)}_\mu$ of mass $m_n$ in 4-dimensional Minkowski 
space and, as mentioned above, for the zero mode (with $m_n = 0$), 
$S_{_A}$ has 4-dimensional gauge invariance.

Here we note that we could have also
derived the above differential equation from examining the $M = \mu$ 
components of the 5-dimensional Maxwell's equation
\begin{equation}
\frac{1}{\sqrt{- G}} \left(\sqrt{- G} \, F^{^{MN}}\right),_{_N} = 0\,,
\label{Maxwell}
\end{equation}
resulting from the action $S_{_A}$ of the full theory in Eq. (\ref{SA}).
Inserting the KK expansion in (\ref{Amu}) into the $M = 4$ 
component of Maxwell's equation yields
\begin{equation} 
\eta^{\mu \nu} \sum_{n = 0}^\infty \partial_\mu A^{(n)}_\nu \frac{d}{d \phi} 
\, \chi^{(n)} = 0\,.   
\label{Max4}
\end{equation}
For $n = 0$, we have $d \chi^{(0)}/d \phi = 0$ and thus a 
4-dimensional condition is not imposed on the zero mode $A^{(0)}_\nu$; this is
consistent with the gauge invariance of the 4-dimensional 
$U(1)$ theory.  However, for the excited modes,  $d \chi^{(n)}/d \phi \neq 0$ 
and hence we must demand 
\begin{equation}
\eta^{\mu \nu} \partial_\mu A^{(n)}_\nu = 0\,,
\label{Vector}
\end{equation}
as required for massive vector particles in 4-dimensional Minkowski space.    

Defining $z_n \equiv (m_n/k) e^\sigma$ 
and  $f^{(n)} \equiv e^{- \sigma} \chi^{(n)}$ we see that Eq. (\ref{Eigen}) 
can be written in the form
\begin{equation}
\left[z_n^2 \frac{d^2}{d z_n^2} + z_n \frac{d}{d z_n} + 
(z_n^2 - 1)\right] f^{(n)} = 0\,,
\label{Besseleq}
\end{equation}
which is the Bessel equation of order 1.  Therefore, the 
solutions for $\chi^{(n)}$ are
\begin{equation} 
\chi^{(n)} = \frac{e^{\sigma}}{N_n} \left[J_1 (z_n) + 
\alpha_n \, Y_1 (z_n)\right]\,,
 \label{Sol}
\end{equation}
where $N_n$ are the wavefunction normalizations, $J_1$ and $Y_1$ are 
Bessel functions of order 1, and $\alpha_n$ are constant coefficients.
Note that this differs from the case of gravitons\cite{dhr}, where the
solutions involved the second order Bessel functions $J_2$ and $Y_2$.
Hermiticity of the differential operator in Eq. (\ref{Eigen}) requires 
that the first derivative of $\chi^{(n)}$ be continuous at the orbifold fixed
points $\phi = 0$ and $\phi = \pm \pi$.  In the limit $e^{-kr_c\pi}\ll 1$,
continuity of $d \chi^{(n)}/d \phi$ at $\phi=0$ yields the relation
\be
\alpha_n  \approx  - \frac{\pi}{2\left[\ln(x_n/2) - k r_c \pi + 
\gamma + 1/2\right]}\,,
\label{alph}
\ee
and at $\phi=\pm\pi$ we obtain the following differential equation
\be
J_1(x_n)  +   x_n J^\prime_1(x_n) + \alpha_n \left[Y_1(x_n) + 
x_n Y^\prime_1(x_n)\right] = 0\,,
\label{xneq}
\ee
where $x_n \equiv (m_n/k) e^{k r_c \pi}$, $\gamma \approx 0.577$ 
is Euler's constant, and 
we have assumed that $m_n \ll k$.  
From these equations, we see that the solutions for $x_n$ depend on the 
value of the model parameter $k r_c$.  To estimate this parameter we note that
the weak scale $\Lambda_\pi$ is
related to $\overline{M}_{Pl}$ by $\Lambda_\pi = \overline{M}_{Pl} \, 
e^{- k r_c \pi}$, and hence to have 
$100~{\rm GeV} < \Lambda_\pi < 1000~{\rm GeV}$, we 
need $11 < k r_c < 12$.  For the low lying modes, varying $k r_c$ 
within this range will not significantly
change the values of $x_n$ (the results are only modified by a few percent)
and for definiteness we take $\Lambda_\pi = 1000$ GeV, 
corresponding to $k r_c \approx 11.27$.  
A numerical solution of Eq. (\ref{xneq}) 
then yields $x_1 \approx 2.45, x_2 \approx 5.57, x_3 \approx 8.70$, and 
$x_4 \approx 11.84$, for the first 4 massive KK modes $A^{(n)}_\mu$ with
$m_n=kx_ne^{-kr_c\pi}$.

It is important to contrast the gauge field KK spectrum with the 
corresponding KK states for gravitons\cite{dhr}.  For gravitons we found
that the KK masses are given by $M_n=k\tilde x_ne^{-kr_c\pi}$, where the
$\tilde x_n$ are roots of the $J_1$ Bessel function, \ie, $J_1(\tilde x_n)=0$,
with $\tilde x_n=3.83\,, 7.02\,, 10.17\,,$ and 13.32 for the first few states.
Comparison of the values of the roots $x_n$ with $\tilde x_n$ shows that
level by level, the KK excitations of the gauge bosons are significantly
lighter than those of the corresponding graviton excitations.

\section{KK Couplings to Fermions}

We now consider the coupling of the gauge KK modes to 
fermions on the 3-brane corresponding to the visible universe.  
The fermion kinetic and gauge interaction terms are given by
\begin{equation}
S_{\psi} = i \int d^4x \int d \phi \left[det (V)\right] 
\overline{\psi} \gamma^\alpha \, V^{^M}_\alpha (\partial_\mu + i g_5 
A_\mu) \psi \, \delta^\mu_{_M} \delta (\phi - \pi)\,,
\label{SfermiI}
\end{equation}
where $V^{^M}_\alpha$ is the vierbein given by
\begin{equation}
G_{_{MN}} = V^\alpha_{_M} \, V^\beta_{_N} \, \eta_{\alpha \beta}    
\label{vierbein}
\end{equation}
with
\begin{equation}       
V^4_4 = 1 \, \, ; \, \, V^\alpha_\mu = e^{- \sigma} 
\delta^\alpha_\mu \, \, ; \, \, det (V) = e^{- 4 \sigma}\,.    
 \label{vierexp}
\end{equation}
Here, $\gamma^\alpha$ are the 
Minkowski space Dirac $\gamma$-matrices, and $g_5$ is the 5-dimensional 
$U(1)$ coupling strength.
Upon integration over $\phi \in [-\pi, \pi]$
and using the KK expansion in (\ref{Amu}), we obtain for the gauge-fermion
interaction term
\begin{equation} 
S_{\psi}^{\rm int} = - \int d^4x g_5 \, \overline{\psi} \gamma^\mu 
\left(\sum_{n = 0}^\infty A_{\mu}^{(n)} (x) \, 
\frac{\chi^{(n)}(\pi)}{\sqrt{r_c}}\right) \psi  \,,
\label{SfermiII}
\end{equation}
where we have employed the redefinition $\psi\to e^{3\sigma (\pi)/2}\,\psi$. 

In order to derive the effective 4-dimensional coupling, we need to 
know the normalization $N_n$ of $\chi^{(n)}(\phi)$.  We note that the 
wavefunction for the zero mode is a constant and that the orthonormality
condition (\ref{Norm}) yields 
\begin{equation}
\chi^{(0)} = \frac{1}{\sqrt{2 \pi}}\,. 
\label{zero}
\end{equation}
For the excited modes with $n \neq 0$, we see that
Eq. (\ref{alph}) gives 
$\alpha_n \sim 10^{- 2}$ for the low lying states.  Thus, within a few 
percent error, the $Y_1$ term, which is proportional to 
$\alpha_n$, can be neglected in the solution for $\chi^{(n)}(\phi)$.  
Using the orthonormality condition we then find   
\begin{equation}
N_n \approx \frac{e^{k r_c \pi}}{\sqrt{k r_c}} \, J_1(x_n)\,.
\label{Nn}
\end{equation}
Defining $g\equiv g_5/\sqrt{2\pi r_c}$, where $g$ is the effective 
4-dimensional $U(1)$ coupling constant, this yields
\begin{equation} 
S_{\psi}^{\rm int} \approx - \int d^4 x g \, \overline{\psi} \gamma^\mu 
\left(A_{\mu}^{(0)} (x) + \sqrt{2 \pi k r_c} \, 
\sum_{n = 1}^\infty A_{\mu}^{(n)} (x)\right) \psi 
\label{SfermiIII}
\end{equation}
for the gauge-fermion interaction term.
Taking  $k r_c \approx 11.27$, we obtain $\sqrt{2 \pi k r_c} 
\approx 8.4$.  Therefore, the excited KK modes couple to the 
3-brane fermions about 8 times more strongly than the 
zero mode, which is identified with the usual `photon' of the 4-dimensional 
Minkowski space.  It is clear that by following the same procedure as
above for the non-abelian gauge fields\cite{sminbulk} we will find that
the KK excitations of all the SM fields are universally more strongly
coupled than the zero mode by the factor $\sqrt{2\pi kr_c}$.
This fact has significant phenomenological 
implications that will be discussed in the next section.

\section{Phenomenological Constraints}

We are now ready to explore the phenomenological consequences of the gauge
KK towers.  In particular, we examine the influence of these KK states
on electroweak precision data, assuming that the KK fields are the only
source of new physics that perturb the SM predictions for these variables. 
In particular, we neglect contributions from graviton exchange as we expect
them to be small.

To begin this analysis, we first realize that the above discussion regarding 
$U(1)$ fields in the RS bulk can be immediately generalized to the case 
of non-Abelian gauge groups as is appropriate for the SM. In particular we 
note that the mass spectra of the excited states of the $W$, $Z$ and $\gamma$ 
towers will be given by the roots of Eq. (\ref{xneq}) 
plus small corrections due to 
the appropriate zero mode masses. In addition, the couplings of all the 
excitations of the SM gauge fields to the fermions on the brane will be 
enhanced relative to their zero 
modes by the same amount, $\sqrt {2\pi kr_c}$. Except for the excitation 
mass spectrum and the precise value of the relative coupling enhancement, we 
see that this situation very closely resembles the physics of the more 
conventional scenario of placing SM gauge fields in the 5-dimensional
bulk of a factorizable 
geometry. Such a scenario has been studied in some detail by many authors in 
order to obtain a bound on the mass of the lightest KK 
state{\cite {sminbulk,getV}}. Below, we follow closely the analysis 
as presented in Ref. {\cite {getV}} but employ the more recent precision 
electroweak data as presented at the summer 1999 conferences{\cite {ewdata}}. 
We assume that even though the gauge field couplings are large, a 
leading order estimate will yield qualitatively correct results. 

We consider the limit where the KK tower exchanges can be described as a
set of contact interactions by integrating out the tower fields.  In this
case, the tower exchanges lead to new dimension-six operators whose
coefficients are proportional to a single fixed dimensionless quantity
\be
V=\sum_{n=1}^\infty {g_n^2\over g_0^2}{M_W^2\over m_n^2} \,.
\ee
Although the couplings are large, we treat $V$ as a small parameter since 
$M_W/m_n$ is small enough to compensate for the couplings.  
The effects of KK exchanges on the electroweak observables, calculated to 
leading order in $V$, are delineated in Ref. \cite{getV}.  These corrections 
include the contributions from tree-level KK interactions and
KK states mixing with the zero modes, in addition
to the usual loop corrections from the zero mode states, or SM fields.  
It is assumed that loop
corrections involving the KK states are higher order and that tree-level
contributions from exchanged KK states can be neglected on the $Z$-pole. A 
second parameter, $s_\phi$, is also required in this analysis to describe 
whether or not the SM Higgs 
field is in the bulk or on the wall. We let this parameter vary over its 
entire allowed range in the analysis below, but as we will see, 
it will have little influence on our final result. 

The electroweak observables used in our global analysis are the leptonic
width of the $Z$, $M_W$, $\sin^2\theta_w^{\rm eff}$ as given by a combined 
determination of all the electroweak asymmetries, $A_b$, $A_c$, 
$R_b$, $R_c$, $Q_W$ - the weak
charge of atomic parity violation, and $\sin^2\theta_w^{\nu N}$ as measured
in deep inelastic neutrino scattering.  Note that at tree-level, graviton
exchange would only contribute to one of these observables, namely
$\sin^2\theta_w^{\nu N}$.  The SM loop corrections involving
the light zero-mode states were computed numerically with 
ZFITTER6.21\cite{zfit}.
Performing a $\chi^2$ fit to the most recent data set\cite{ewdata} and
assuming only that the Higgs boson mass is $\geq 100$ GeV\cite{ewdata} yields
the constraint
\be
V\leq 0.0010-0.0013
\ee
at $95\%$ C.L, where the range results from varying the parameter $s_\phi$. 
We simply assume the weaker bound, $V<0.0013$, in what follows. We note that 
this bound allows for variations in 
both the input values of the top quark mass, $\alpha(M_Z)$, and 
$\alpha_s(M_Z)$, as well as systematic 
effects as described in {\cite {getV}}. 

Given the ratio of coupling strengths derived in the above
section, \ie, $g_n/g_0=\sqrt{2\pi kr_c}\approx 8.4$, and using
\be
\sum_{n=1}^\infty {x_1^2\over x_n^2}\approx 1.5\,,
\ee
implies that the
mass of the first gauge boson excited state is bounded by $m_1\gsim 23$ TeV.  
It is interesting to note that this bound implies a corresponding constraint
of $M_1\gsim 36$ TeV on the mass of the first KK graviton resonance.  Since
both of these lower bounds on the first excitation mass are about a factor
of 100 or more larger than the SM Higgs vacuum expectation value, one may
worry that we are in danger of forming another hierarchy.
Since $m_n=kx_ne^{-kr_c\pi}$, with $x_n$ given above, this yields the 
constraint $ke^{-kr_c\pi}\gsim 9.4$ TeV. Taking the conservative value 
$\Lambda_\pi=1$ TeV for the weak scale and folding in the 
explicit definition of $\Lambda_\pi$ as well as
the relationship in Eq. (\ref{mpl}), we finally arrive at the constraint
on the RS model parameters of
\be
{k\over M} \gsim 4.5 \,.
\ee
This implies that the magnitude of the bulk curvature violates the initial
assumption of the theory that $|R_5|=20k^2<M^2$.   In turn, if we demand that
this assumption holds, then the weak scale is forced to be very large with
$\Lambda_\pi\gsim 100$ TeV.  Note that if we had taken a 
smaller value for $\Lambda_\pi$ and/or the tighter constraint on $V$ 
the above bound (27) on this ratio of RS parameters would have been stronger by 
as much as a factor of 3. 

\section{Conclusions}

In this paper we have explored the phenomenological viability of placing
gauge fields in the bulk of the Randall-Sundrum model of localized gravity.
We derived the gauge field KK spectrum from examination of the action of
the theory and also from analyzing the 5-dimensional Maxwell's equation.
We then computed the gauge-fermion interactions on the SM 3-brane and found
that the excited KK states couple $\sim 8$ times more strongly than the 
zero-modes.  The influence of these strongly-coupled gauge KK states on
electroweak precision data was investigated with the resulting constraint on
the mass of the first excited state of $m_1\gsim 23$ TeV.  Assuming 
$\Lambda_\pi\sim 1$ TeV, this in turn
implies a bound on the model parameters of $k/M\gsim 4.5$, which suggests
that the bulk curvature is too large by a factor of $\sim 20$ to trust the 
RS metric (\ref{5metric}) as a solution to Einstein's equations.  The weak
scale must be pushed to $\Lambda_\pi\gsim 100$ TeV in order to preserve
$|R_5|\lsim M^2$.
Hence, as a solution to the hierarchy problem,
the model as presently formulated is inconsistent with gauge fields
existing in the bulk.  The effects of higher order curvature terms
must be examined in order to determine the robustness of the theory.

\noindent{\bf Acknowledgements}:

We would like to thank Y. Grossman, M. Schmaltz, R. Sundrum, M. Wise 
for beneficial discussions.  H.D. acknowledges discussions with L. Dixon
and M. Peskin.

%
\def\IJMP #1 #2 #3 {Int. J. Mod. Phys. A {\bf#1},\ #2 (#3)}
\def\MPL #1 #2 #3 {Mod. Phys. Lett. A {\bf#1},\ #2 (#3)}
\def\NPB #1 #2 #3 {Nucl. Phys. {\bf#1},\ #2 (#3)}
\def\PLBold #1 #2 #3 {Phys. Lett. {\bf#1},\ #2 (#3)}
\def\PLB #1 #2 #3 {Phys. Lett.  {\bf#1},\ #2 (#3)}
\def\PR #1 #2 #3 {Phys. Rep. {\bf#1},\ #2 (#3)}
\def\PRD #1 #2 #3 {Phys. Rev.  {\bf#1},\ #2 (#3)}
\def\PRL #1 #2 #3 {Phys. Rev. Lett. {\bf#1},\ #2 (#3)}
\def\PTT #1 #2 #3 {Prog. Theor. Phys. {\bf#1},\ #2 (#3)}
\def\RMP #1 #2 #3 {Rev. Mod. Phys. {\bf#1},\ #2 (#3)}
\def\ZPC #1 #2 #3 {Z. Phys. C {\bf#1},\ #2 (#3)}

\end{document}